\documentclass[conference]{IEEEtran}
\IEEEoverridecommandlockouts
\usepackage{amsmath}
\usepackage{amsthm}
\usepackage{graphicx}
\usepackage{color}
\usepackage{cite}
\usepackage{enumerate}
\begin{document}
\title{Introduction of Improved Repairing Locality into Secret Sharing Schemes with Perfect Security}
\author{
\IEEEauthorblockN{Yue Fu, Shuhao Sun, Dagang Li}
\IEEEauthorblockA{School of Electronic and Computer Engineering\\
Institute of Big Data Technologies\\
Peking University\\
Shenzhen, China, P.R.\\
Email: fuyuefyu@126.com, dgli@pkusz.edu.cn}
\and
\IEEEauthorblockN{Peng Liu}
\IEEEauthorblockA{School of Computer Science and Technology\\
Hangzhou Dianzi University\\
Hangzhou, China, P.R.}
\thanks{Corresponding author: Dagang Li. This work was supported by Shenzhen Research Program JCYJ20150626110611869, ZDSYS201603311739428 and the Shenzhen Municipal Development and Reform Commission (Disciplinary Development Program for Data Science and Intelligent Computing).}
}
\maketitle
\begin{abstract}Repairing locality is an appreciated feature for distributed storage, in which a damaged or lost data share can be repaired by accessing a subset of other shares much smaller than is required for decoding the complete data. However for Secret Sharing (SS) schemes, it has been proven theoretically that local repairing can not be achieved with perfect security for the majority of threshold SS schemes, where all the shares are equally regarded in both secret recovering and share repairing.
In this paper we make an attempt on decoupling the two processes to make secure local repairing possible. Dedicated repairing redundancies only for the repairing process are generated, which are random numbers to the original secret. Through this manner a threshold SS scheme with improved repairing locality is achieved on the condition that security of repairing redundancies is ensured, or else our scheme degenerates into a perfect access structure that is equivalent to the best existing schemes can do. To maximize security of the repairing redundancies, a random placement mechanism is also proposed.
\end{abstract}
\section{Introduction}
\subsection{Backgrounds}$(k,n)$ threshold Secret Sharing (SS) scheme was first proposed by Shamir\cite{1} and Blakley\cite{2} independently in 1979. With Secret Sharing, a secret can only be recovered if threshold $k$ number of shares from the total $n$ is reached. Any number of shares less than $k$ reveals zero information about the secret so security is well-guaranteed. The difference between $n-k$ can be regarded as some safeguard against share failures, so the secret will not be jeopardized should one or two shares fail. However in practical systems, failed shares should be repaired or replaced in time to maintain a comfortable $n-k$ margin. In most cases a failed share will be repaired with the help of other participants, preferably those in the neighborhood so minimum network resources are involved in the process.\\
\indent The most simple way of repairing a share is to first recover the secret and then generate a new share with it. However this process requires accessing at least a threshold of shares and the revealing of the secret itself. To solve these issues, existing repairable SS schemes can be divided into two categories. The repairability of the first category is based on enrollment protocols\cite{3,4,5,6,7,8,9,10,11,12} while the other is based on combining SS with regenerating codes. Enrollment protocols are used to repair a secret share by only using coded version of other shares, so the shares and the secret themselves are not revealed. However, these enrollment protocols can only be applied to specific SS schemes so the generality is lost.\\
\indent Dimakis et. al. applied network coding to distributed storage system and proposed the concept of regenerating codes then analysed the trade-off between the cost of storage and bandwidth\cite{13}. In recent years, the combination of SS and regenerating codes appears. They make any failed share repairable via cooperation of no less than $d$ participants, where $d$ is called repairing degree. In 2014, Guang et. al. pointed out that a naive combination may lead to loss of perfect security and proposed a GLF scheme as a solution without loss of perfect security\cite{14}. Rawat et. al. proposed a centralized multi-node repairing scheme, in which MBR and MDR codes were discussed\cite{15}. In 2016, Stinson et. al. proposed another solution based on ramp scheme, which achieved the ``restricted repairing" that the authors defined, then compared it with the GLF scheme.\\
\indent However, in these schemes, repairing degree $d$ should be no less than threshold $k$: according to their definitions, when $d<k$, since any set of $d$ participants can repair any number of failed shares so effectively secret-recovery threshold $k$ is lowered to $d$. To compromise the conflict between repairing mechanism and threshold in the context, most existing schemes assume that $d>k$. Better efficiency in repairing is achieved by requiring only a smaller amount of data from each of the $d$ nodes so the total amount is reduced.\\
\indent In the repairing process, we may also hope repair to be done by accessing a smaller than $k$ number of other nodes, because shares can be widely distributed and accessing nodes far away can also be very costly. The notion of locally repairable codes\cite{16} can be extended into SS schemes. In a code with locality $r$, any symbol of a codeword can be deduced by at most $r$ other symbols of the codeword. Unfortunately, it was proven theoretically that a threshold SS scheme is not locally repairable. Still, there exists perfect access structures with small $r$\cite{17}. This means under the threshold paradigm, introduction of erasure codes leads to loss of locality or the threshold scheme turns out to be a general access structure.\\
\indent The main purpose of this paper is to provide new mechanisms for repairing secret shares while satisfying perfect security under some conditions where the repair degree $d$ less than SS threshold $k$. We make an attempt on decoupling the two processes by setting dedicated repairing redundancies only for repairing process. Then IDA is adopted to make the data separated: one part is ``secret recovery only" while the other is ``repairing only". Then we get a weak result: When security of repairing data is ensured, a threshold SS scheme with improved repairing locality is achieved, or else our scheme degenerates into a perfect access structure, which is equivalent to what locally repairable codes can do. Finally, a random placement mechanism is proposed to maximize security of repairing data so this condition can be better satisfied.\\

\subsection{Contributions}The contributions of this paper are summarized as below:
\begin{enumerate}
\item \textbf{Repairing Locality.} In existing scenarios, maintenance of threshold results in loss of repairing locality. In our mechanism the repairing process is bounded inside node groups so locality of repairing process is achieved, which can reduce the repairing degree $d$ much lower than the SS threshold $k$.
\item \textbf{Decoupled Repairing Process.} This paper proposes a weak condition (security of repairing data) such that:\\
(1) When security of repairing data is ensured, the local repairing process relies on redundancies that are decoupled from the original SS so a threshold SS scheme with improved repairing locality is achieved.\\ 
(2) When the repairing data security is lost, the security level of our scheme is no worse than the best existing schemes (equivalent to a perfect access structure with locally repairable codes).
\item \textbf{Redundancy Distribution Protocol.} For the sake of maximizing security of repairing data (the weak condition in last item), a random placement mechanism is proposed to hide the redundancy shares during the distribution in IDA process.
\end{enumerate}
\subsection{Paper outline.}The rest of the paper is organized as follows. In section \uppercase\expandafter{\romannumeral2}, some preliminaries and related work will be introduced briefly. In section \uppercase\expandafter{\romannumeral3}, a localized repairable grouped SS scheme based on grouping scheme and repairing function along with 2 toy models are constructed. In section \uppercase\expandafter{\romannumeral4}, parameters in our model will be discussed in details. In section \uppercase\expandafter{\romannumeral5}, properties will be discussed and comparison with regenerating codes will be made. Finally, the conclusion will be drawn in section \uppercase\expandafter{\romannumeral6}.
\section{Preliminaries and Related Work}
\subsection{Secret Sharing}
\subsubsection{Backgrounds}$(k,n)$ threshold SS scheme was first proposed by Shamir\cite{1} and Blakley\cite{2} independently in 1979. Afterwards, various SS scheme based on different mathematical models are proposed: SS scheme based on Chinese Remainder Theorem(CRT) by Asmuth-Bloom\cite{20}; SS scheme based on matrix multiplication by Karnin et. al.\cite{21}Among those schemes Shamir's $(k,n)$ threshold scheme is the most commonly used one due to its manifest representation and perfect security.
\subsubsection{Ramp Scheme}In 1985, ramp scheme was proposed by Blakley et. al\cite{22}. It possesses two thresholds $k_{1}<k_{2}$ such that:
\begin{enumerate}[(1)]
\item The secret $s$ can be recovered with the cooperation of no less than $k_{2}$ participants.
\item No information should be revealed about the secret $s$ under the cooperation of less than $k_{1}$ participants so perfect security is achieved.
\item When the number of collected secret shares is between $k_{1}$ and $k_{2}$, the revealed information increases linearly with the increase of the former.
\end{enumerate}

\indent When $k_{1}=k_{2}-1$, a ramp scheme turns out to be a threshold scheme. In 2012, Kurihara et. al. proposed a perfect secure parameter $m$ to build a new regenerating code then proved the security of that scheme is equivalent to a ramp scheme\cite{23}.
\subsection{Repairable Secret Sharing Schemes}In 1998, the first repairing mechanism based on enrollment protocol was proposed by Herzberg et. al.\cite{24} This protocol is based on Shamir SS scheme. The spirit is: make $k$ surviving participants add some specific polynomials to their secret shares so they are not revealed then send the addition to the share-loser. The loser can recover his share via Lagrange interpolation. Still, these schemes will cause too much cost in bandwidth and the loss of generality is also concerned.\\
\indent A type of enrollment protocols can be applied to repairing process in SS. For instance,\cite{3,4,5,6} proposed an enrollment protocol based on Shamir scheme, which is similar to that proposed by Herzberg et. al.\cite{7} An enrollment protocol based on public-verifiable SS schemes is proposed in\cite{8,9}, which introduced public-verifiable functionality based on \cite{10}. Moreover, Saxena et. al. proposed enrollment protocol based on bivariate polynomial SS\cite{11}. Yue et. al. proposed enrollment protocol based on vector space SS\cite{12}.\\
\indent Still, these enrollment schemes mostly depend on specific SS schemes so they are poor in generality, and the repairing still needs $k$ nodes. Hence, new repairable SS schemes without loss of generality and universality is still missing.
\subsection{Regenerating Codes}Dimakis et. al. applied network coding to distributed storage system and proposed the concept of regenerating codes then analysed the trade-off between the cost of storage and bandwidth\cite{13}. The principle of $(n,k,d,\alpha,\beta)$is: code the original files then save them into $n$ nodes. Each of the nodes keeps data of size $\alpha$. When any of these nodes fails, the system repairs it via downloading data from $d$ surviving nodes. Each of them offers data of size $\beta$. Meanwhile, any $k$ participants should be able to recover the original file. In recent years, several construction of regenerating codes were proposed\cite{25,26}.
\subsection{Locally Repairable Codes}Locally repairable code is proposed to trade more storage space for lower repairing bandwidth. The number of nodes to be accessed during repairing process is lower than the case of other erasure codes. Rawat et. al. proposed a specific locally repairable code scheme in 2014\cite{27}. In their paper, they deeply discussed repairing locality and grouping schemes. Shahabinejad et. al. discussed a binary locally repairable code to realize fast repairing functionality for several nodes in failure\cite{28,29}. \cite{30}optimized LRC based on maximum rank distance Gabidulin codes. \cite{31}considered locally repairable codes over small fields and proposed new constructions of optimal binary and q-ary cyclic (and linear) codes with locality and availability.\\

\section{A Localized Repairable Grouped SS Scheme based on Grouping Scheme and Repairing Function}Our scheme is based on grouping the secret shares. For the sake of locality, the participants of the original SS are divided into disjoint groups, preferably in a compatible way to the access structure or some multilevel SS property.\\
\indent In practice, the grouping may need to consider multiple factors such as physical location of the servers or different ranks in the multilevel cases. Sometimes these factors may conflict with each other and if that occurs, we should treat the problem thoughtfully to achieve the best overall grouping result. For example, in multilevel SS schemes, the grouping should comply more with the level of participants than the distance among them.\\
\indent Moreover, the grouping should have no affect on their role and functionality in the SS scheme. Or in other words, the original SS won't see this grouping and treat all shares as they originally should be. This means that the shares in each group should keep as they are and cannot be re-encoded for the repairing capability.
\subsection{Construction}Now, we will build our model on the base of these SS groups and repairing functions. Firstly, a repairing function is introduced inside each group to make the repairing local and secure without modifying the secret shares. The participants of the group constitute a repairing function and a repairing redundancy to keep information of that function is generated. The function should contain all information to repair the secret shares within the group. The local repairing information are stored through the generated repairing function of each group. It can be defined as: given a set of points $(x_{i},y_{i})$, a repairing function of these point is a function $y=f(x)$ such that $y_{i}=f(x_{i})$.\\
\indent Furthermore, saving all the coefficients of that polynomial or saving only a determining point for the polynomial is also a consideration. So we introduce the notion of strong repairing redundancy and weak repairing redundancy to measure the significance of the repairing redundancy:\\
\indent \textbf{\emph{Strong repairing redundancy.}} A redundancy is called \textit{strong repairing redundancy} if it can determine the whole repairing function independently. For example in the following toy model, the combination of all the coefficients of the polynomial is a strong repairing redundancy.\\
\indent \textbf{\emph{Weak repairing redundancy.}} A redundancy is called \textit{weak repairing redundancy} if it can determine the repairing function with the cooperation from others. For example in the following toy model, the redundant point of the polynomial at $P_{\lambda}$ is a weak repairing redundancy.\\
\indent For the convenience of understanding, a toy model based on polynomial repairing function is given in the following context.
\subsubsection{Difference between trivial Shamir SS scheme and the one used in our model}In Shamir scheme, secret shares are in the form of $(x_{i},y_{i})$. We can either save the pair as the secret share or only $y_{i}$ and make $x_{i}$ public. Making $x_{i}$ public will not affect the secrecy because it is the polynomial who determines the secret, and without the corresponding $y_{i}$, the knowledge of $x_{i}$ reveals zero information about the secret. However, to make local repair possible inside a group we cannot save the $(x_{i},y_{i})$ pair together, because in that case when the share fails we lose both $x_{i}$ and $y_{i}$. For Shamir SS there are still enough shares to recover the polynomial and the secret, so it can choose a new random $x_{i}'$ and the corresponding $y_{i}'$ to replace the failed one. However the shares from one group are not sufficient to recover the polynomial for the secret, so the local repair has to find the exact $(x_{i},y_{i})$ of the failed share back, or the global SS will fall apart. In our repairing function, $x_{i}$ can be the identification of the storage server or some other common information. The publicity can be carried out in two ways: one is to publish every $x_{i}$ to every participants while the other is to put $x_{i}$s to some common place.\\
\emph{\textbf{\indent Publication of $x_{i}$.}} As discussed before, in our model we don't want $x_{i}$s to be part of the secret shares so we should publish $x_{i}$. In practical applications, the storage servers are sometimes equipped with well-defined identity numbers due to some design or operational considerations, and these numbers can be used as $x_{i}$ for the $i$-th server. If such easy solution is not possible, we need to have our own publishing mechanism.\\
\indent A trivial consideration is setting a dedicated server to offer services about renewing and maintaining these numbers for the servers. When any server is down or replaced, the dedicated server should give the replacement server the correct sequence number $x_{i}$. Of course when the number of storage servers becomes huge, keeping all $x_{i}$ at one server may become an issue. In that case we may want to store the $x_{i}$s of the group members inside the group.\\
\indent Furthermore, that dedicated server becomes a single-point-of-failure of the whole repairing mechanism. To relief that problem we can put the $x_{i}$s of the group members together with the repairing function as the repairing redundancies and distribute it to group members.\\
\indent The publication of $x_{i}$s is the first step of our model.
\subsubsection{A toy model of grouped SS with repairing function}Firstly, we build a normal $(k,n)$ SS, let's say $(8,12)$, and divide the 12 shares into 3 groups with 4 shares each. In each group, a local repairing function is generated that is only available to that group, which helps repair failed shares of the group. Let's consider group 1 whose participants are $(P_{1},P_{2},P_{3},P_{4})$ with secret shares $((x_{1},y_{1}),(x_{2},y_{2}),(x_{3},y_{3}),(x_{4},y_{4}))$. The process is summarized in the following steps.\\
\indent \emph{Step 1:} Generate a 3-degree polynomial with coefficients $(b_{1},b_{2},b_{3},b_{4})$. Determine their values with the points at $((x_{1},y_{1}),(x_{2},y_{2}),(x_{3},y_{3}),(x_{4},y_{4}))$.\\
\indent \emph{Step 2:} Choose a random $x_{\lambda}$ distinct to existing $x_{i}$s. Get the corresponding $y_{\lambda}$ from the polynomial. Note here the generated $y_{\lambda}$ is just a random point to the global secret S. As shown in Fig. 1.\\
\begin{figure}[!t]
  \centering
  \includegraphics[width=3.5in]{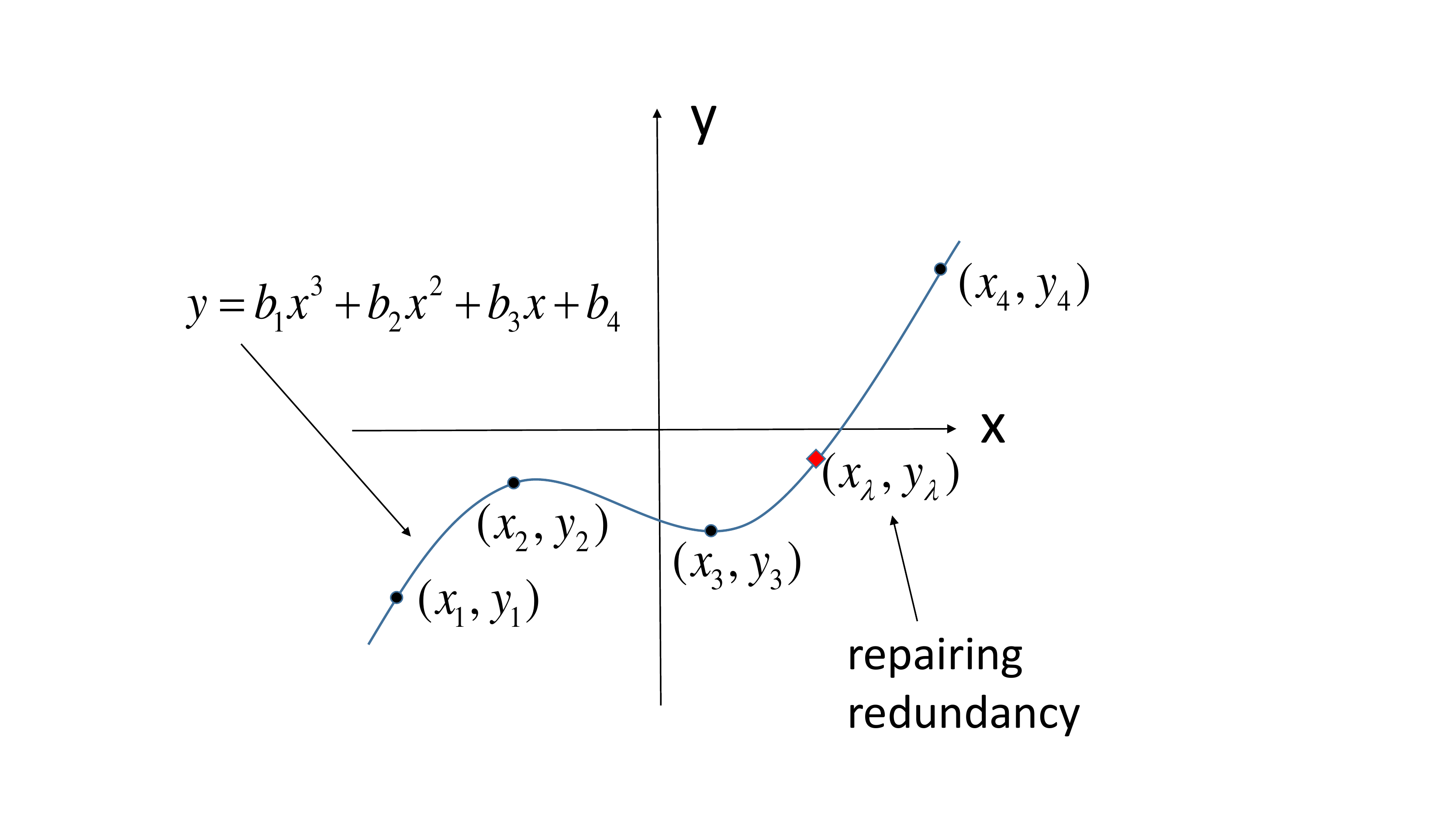}
  \caption{A repairing function for $((x_{1},y_{1}),(x_{2},y_{2}),(x_{3},y_{3}),(x_{4},y_{4}))$ is a polynomial passes through these points with degree 3. $(x_{\lambda},y_{\lambda})$ is a random point on that curve.}
\end{figure}
\indent \emph{Step 3:} Treat $(b_{1},b_{2},b_{3},b_{4})$ or $y_{\lambda}$ as the sub-secret of the group and save it to a reliable server inside group 1 that we call $P_{\lambda}$.\\
\indent \emph{\textbf{Analysis:}} When any of the shares at $(P_{1},P_{2},P_{3},P_{4})$ is missing, we can access $P_{\lambda}$ to recover the missing share. However, this will cause extra cost in hardware dedicated just for repair. On the other hand, $P_{\lambda}$ can play a largely different role with regard to the original shares of the group. If we store $(b_{1},b_{2},b_{3},b_{4})$ in it, anyone with the public $x_{i}(i\in \lbrace 1,2,3,4 \rbrace)$ can get $y_{i}$ from $P_{\lambda}$, which means $P_{\lambda}$ effectively possesses all shares of the group. In this case, $P_{\lambda}$ is a strong repairing redundancy that needs careful treatment. If we just store $y_{\lambda}$ in it, group 1 simply becomes a local $(4,5)$ threshold scheme for the sub-secret. In this case, $y_{\lambda}$ is a weak redundancy. When the repairing redundancy is too abundant, it may be a clear target for attack and weak point for security. Still, saving too much data outside the group leads to loss of locality in repairing process.\\
\indent We suggest Information Distribution Algorithms (IDA) to get rid of the dedicated servers and hide the repairing redundancies deeper inside the system. It should be carefully chosen according to the specific repairing redundancy. For example, Shamir's scheme maybe too much of a overkill for weak redundancy, for which we can use other more cost-efficient SS schemes to reduce the cost to match with its significance, for example with the SS-made-short scheme proposed by H. Krawczyk\cite{18}.\\
\indent In the following part we will show Second Secret Sharing (SSS) as a toy model of IDA for weak redundancy.
\begin{figure}[!t]
  \centering
  \includegraphics[width=3.5in]{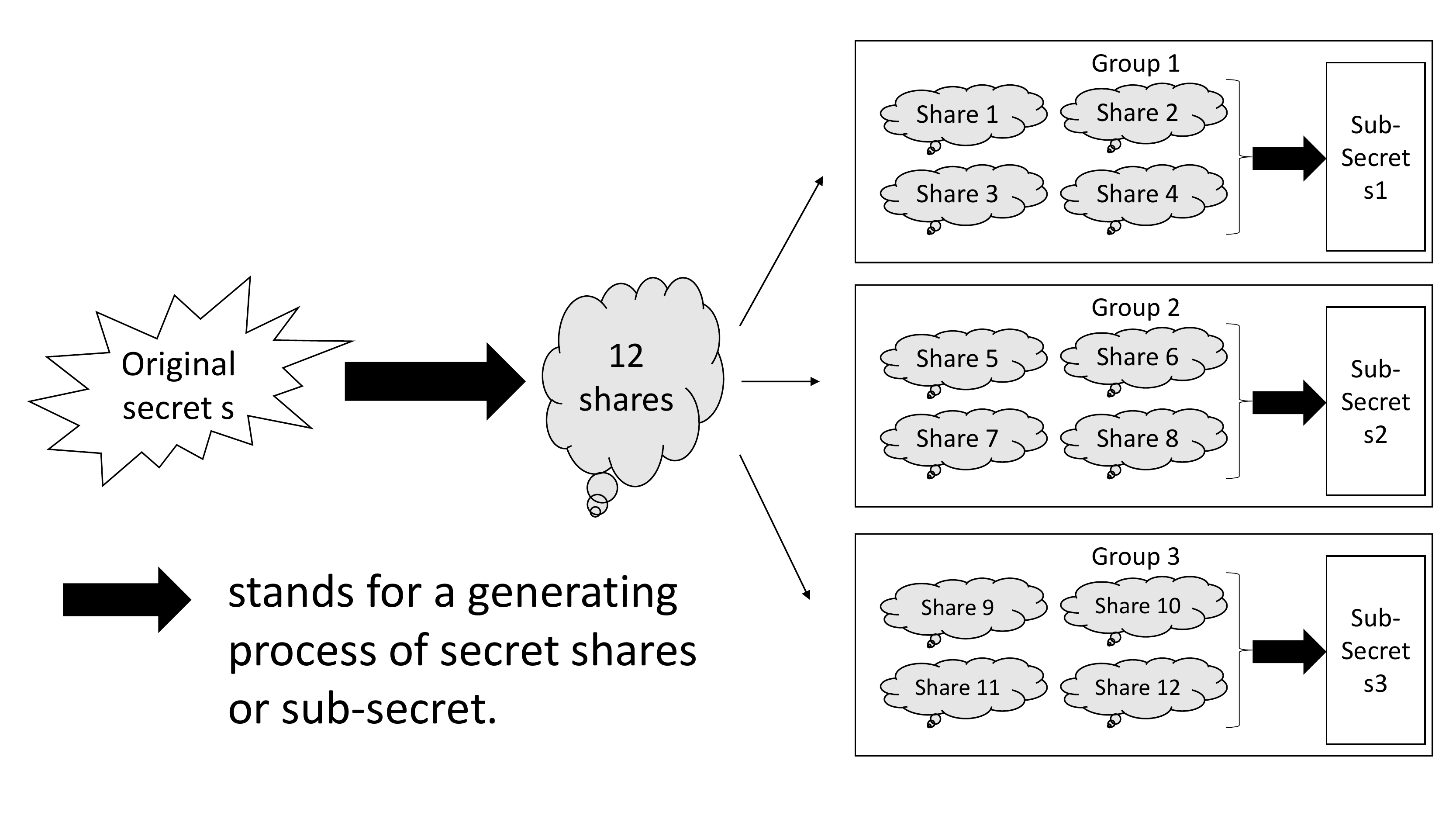}
  \caption{Process of grouped SS along with repairing function}
  \label{fig_rf}
\end{figure}
\subsubsection{Grouped SS with Second Secret Sharing}
\begin{figure}[!t]
  \centering
  \includegraphics[width=3.5in]{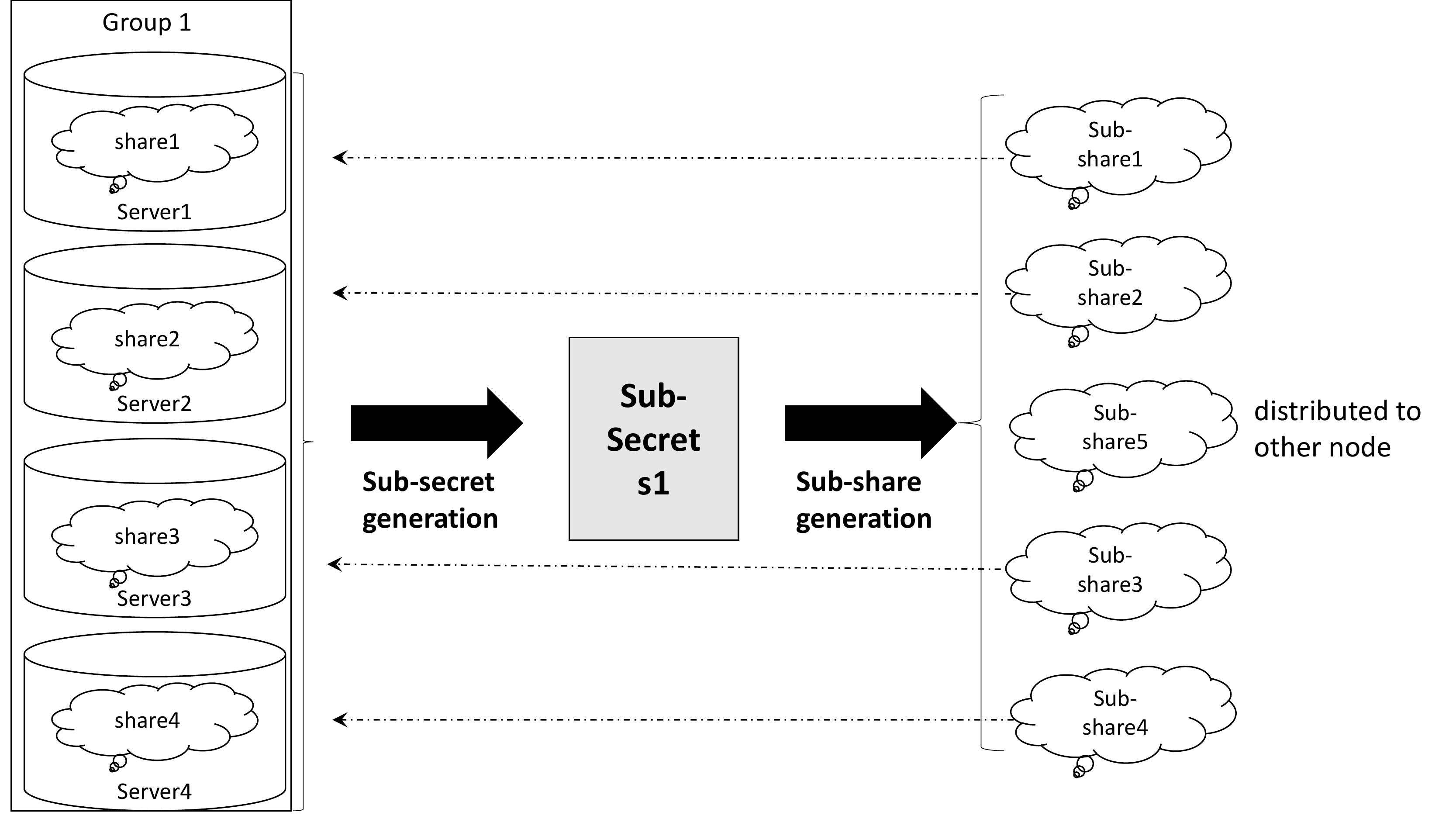}
  \caption{SSS in Group 1 as an example}
  \label{fig_sss}
\end{figure}

We propose our final model as a localized repairable grouped SS model armed with SSS method. In this model, weak repairing redundancies are chosen to be the sub-secrets in order to recover the repairing function. SS is used a second time as IDA to distribute weak redundancy. It is organized in the following steps:
\begin{enumerate}
\item Consider a $(k,n)$ threshold SS, divide the $n$ participants $(P_{1},P_{2},...,P_{n})$ into $m$ disjoint groups $(G_{1},G_{2},...,G_{m})$ to form a $(m,n)$ grouped SS scheme. Each $P_{i}$ keeps a share of $(x_{i},y_{i})$.
\item Suppose the number of members in group $G_{i}$ to be $N_{i}$. Generate a $(N_{i}-1)$-degree polynomial based on the points determined by the secret shares of the group. As shown in Fig.\ref{fig_rf}.
\item Choose another random point on that polynomial. Treat this random point as the sub-secret.
\item Perform a SSS to share the sub-secret to group members. As shown in Fig.\ref{fig_sss}.
\end{enumerate}

\subsection{Further Design about SSS}
\subsubsection{Security/locality trade-off in SSS process}\emph{\textbf{\\Example.}} As an example, we use the same $(8,12)$ toy model from last subsection, which is grouped into $(G_{1},G_{2},G_{3})$ with 4 shares each. For $G_{1}$ with participants $(P_{1},P_{2},P_{3},P_{4})$, a $y_{\lambda}$ is generated with the steps just mentioned. Next, the $y_{\lambda}$ is treated as the sub-secret $s_{1}$ of the group and another Shamir SS scheme is used to distribute $s_{1}$ to $G_{1}$ members.\\
\indent The threshold of this sub-secret distribution can not be lower than 4, or otherwise it will affect the threshold of the original SS: for example with a (3,4) scheme, when any 3 out of 4 group members are revealed to an attacker, $y_{1}$ and consequently the repairing function can be recovered, with which the attacker can figure out the 4th share by himself.\\
\indent Here in this paper a (4,5) scheme is recommended, therefore we need a server outside the group to put the extra sub-share on. However the location of that extra sub-share should be hidden in some way to avoid abuse of repairing mechanism from attackers who own intact data from a subset of nodes. As shown in Fig. 4, here we are proposing a \textit{redundancy distribution protocol} of random placement to maximize the security:
\begin{enumerate}
\item Assume that each node owns a unique sign, perfectly comes from hardware to verify its identity, for example a MAC address. When any data fails, it propose a repairing request and recover the lost data. Repairing request of any node must be proposed by the server who loses its data, cooperating with the 3 other nodes. This means a repairing request should be authorized by every member of that group.
\item Choose a random server outside the group and store the sub-share there, and give a hash identity calculated with combination of MAC addresses form all 4 nodes in the group.
\item The group doesn't remember the chosen server, and the latter doesn't know which group the sub-share belongs to, so true randomness is achieved.
\item During a repairing process the node who claims a repairing service requests for the sub-share by broadcasting the hash identity, and the server who passes the identity check sends the extra sub-share back to the group.
\item The missing secret share is repaired through repairing function and only revealed to the proposer.
\end{enumerate}

\begin{figure}[!t]
  \centering
  \includegraphics[width=3.5in]{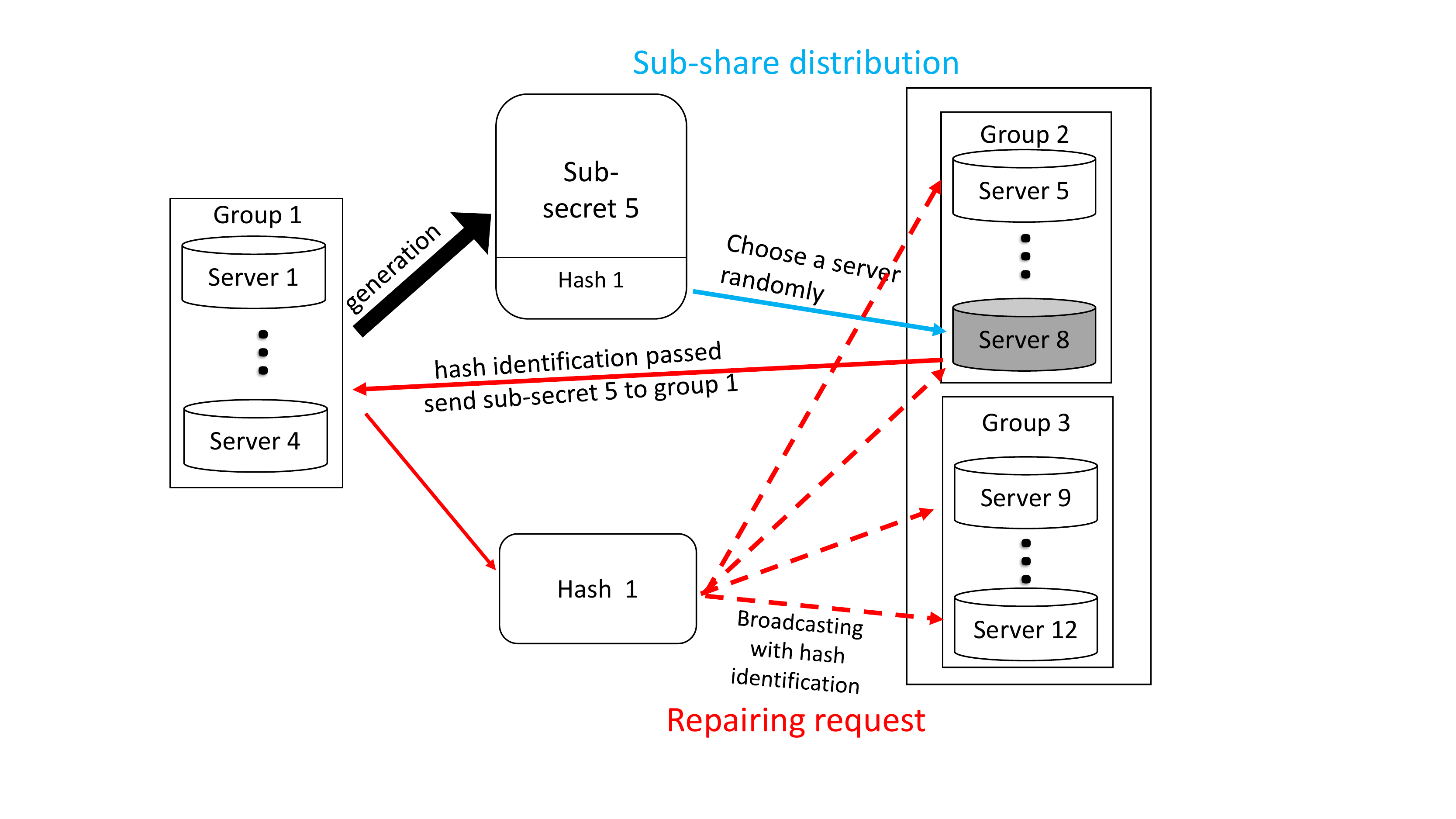}
  \caption{Redundancy Distribution Protocol}
\end{figure}

\indent This design is a trade-off between locality and security. With the protocol proposed, extra shares should be stored out of the group, so improvement of security leads to tiny loss of locality.
\section{Further Discussion about our model}
\subsection{Repairing Function}As we see, a repairing function is a map from published $x_{i}$'s to the value of secret shares. Then the ``secret" to be shared turns out to be a function rather than a value so it is essentially a function sharing process. In our model, currently a polynomial function is used which is completely determined by its coefficients. Thus, the repairing redundancy turns out to be a multi-SS process.\\
\indent The polynomial repairing function is somehow similar to the repairing process of SS as discussed by Hayashi et. al.\cite{19}, which repairs or regenerates a broken share by first recovering the original secret along with the secret-distributing polynomial from enough surviving shares. With the knowledge of that polynomial any amount of shares can be repaired or regenerated. However, our repairing function is actually very different from this SS repairing mechanism because the purpose is quite the opposite to each other. The goal of SS is to generate a set of secret shares for a known secret $s$ in hand while the goal of the repairing function (polynomial in this case) is to generate repairing information from existing shares that are already known and then create repairing redundancies based on this function. Thus, in repairing function, what is important is not the new ``secret" ($a_{0}$ in the new polynomial) but the polynomial, which can be regarded essentially as a multi-SS process.
\subsection{Repairing Redundancies}The concept of repairing redundancy is introduced to describe the significance of a redundancy. In the polynomial case, the coefficient $b_i$'s and the share $y_i$'s are elements of the same finite field so they are of the same size. In this sense, a share or a redundancy is equivalent to a coefficient. Thus, the significance of the generated redundancy can be measured by either how few cooperators it relies on to recover the sub-secret or how many coefficients it can independently determine.\\
\indent For example in the toy model of repairing function, the weak redundancy is as significant as 1 share while the strong redundancy is as significant as all 4 shares. Clearly, using a strong redundancy is not wise now because it is strictly worse than a simple replication scenario: the size is the same but more computation is involved. This is one reason why we choose weak redundancy in our model.
\subsection{Decoupled Properties}Our repairing scheme is also different from existing repairable SS schemes that make use of erasure codes such as Reed-Solomon codes and regenerating codes, where the redundancies are simply liner combination of the original shares and therefore of the same global significance. These redundancies should be handled carefully because they may affect the properties of the original SS process.\\
\indent In our scheme the secret recovery process and the share repairing process are decoupled in the sense that the repairing functions are local with no global significance to the original SS process. The repairing functions are the ``secret" of grouped shares that works in the repairing process while the secret shares of the original SS work in the secret recovery process. The data repairing process is independent from secret shares so secret-recovering and share repairing process are decoupled.
\section{Model Analysis}
\label{sec_analysis}
\subsection{Attack/Threat Scenarios}
\subsubsection{Repairing Function Generation}In this process, under the weak redundancy case, the generated local redundancy is kept in an additional server in each group. We will first discuss its resistance under attacks.\\
\indent Let's consider a (8,12) threshold scheme divided into 3 groups with 4 shares each. From each group an attacker can collect 4 shares at most (direct or calculated through repairing) even when he compromise all 5 servers. Hence, we consider how difficult it is for an attacker to get all 4 shares of one group measured by the probability $p$.\\
\indent Suppose all servers behave equally under attack and let us say the probability the attacker gets all data on one server is $q$. Hence, when there is no redundancy at all, the probability $p_{1}$ turns out to be:
\begin{equation}
p_{1}=q^{4}
\end{equation}
\indent In a (4,5) scheme, any 4 out 5 servers can provide all 4 shares of this group, the probability $p_{2}$ turns out to be:
\begin{equation}
p_{2}=C_5^4q^{4}(1-q)+q^{5}=q^{4}(5-4q)
\end{equation}
\begin{equation}
p_{2}-p_{1}=4q^{4}(1-q)
\end{equation}
\indent Since $0\leq q\leq 1$, $p_{2}-p_{1}\geq 0$.\\
\indent Hence inside one group, we've proven that the introduction of new server increases the successful rate for attacks, because the additional server increases the number of targets. SSS is proposed to solve these problems. 
\subsubsection{SSS Process}We've discussed the decoupled property about SSS process in the last section. In this part we will analyze the impact on the secret recovery data when repairing data is under attack.\\
\indent Let's quote the $(8,12)$ scheme as an example again and consider secret recovery data. When data from compromised servers are not sufficient to carry out any local repair in repairing data, the secret recovery threshold 8 is well-retained; otherwise additional shares can be obtained via the repairing process, which effectively lower the threshold. In the worst case minimally 6 compromised servers can reveal the original secret.\\
\indent In the worst case, the attacker gets 6 shares from only two groups with 3 from each, and unfortunately for each of the two groups the repair redundancy is stored among the compromised servers of the other group. It is easy to see that a fourth share can be ``repaired" from each group, so in total the attacker has 8 shares at hand which is sufficient to recover the original secret.\\
\indent Similarly, if the data on the compromised servers are sufficient only for just one repairing, then the attacker needs to break minimally 7 servers. Actually we can increase the number of required servers from 6 to 7 for the worst case scenario, if we never store the redundancy to each other between any 2 groups, which is at the cost of less randomness in the redundancy distribution protocol.\\
\indent All in all, the threshold scheme can be retained when data from compromised servers can not support any repair. The security of the redundancies in repairing data is protected by the randomness protocol. When repairing data is no longer secure, the secret recovery threshold breaks up and it turns out to be an access structure with perfect security. In any case, our model provides support to repairing locality so the condition that repairing degree $d<$ threshold $k$ is no longer required.
\subsection{Comparison with Regenerating Code}In this part, the performance of our model is compared with repairing mechanism based on regenerating codes.
\subsubsection{Storage}In our model, repairing redundancies are generated and distributed by IDA. We've shown SSS as an example. In the following discussion, we are calculating the storage overhead in repairing redundancy generation and SSS process.\\
\emph{\indent Repairing Redundancy Generation.} Assume that the capacity of original secret s to be 1. During the SSS process, we have m groups with m sub-secrets $s_{i}$ in total. In this process, the additional capacity turns out to be m.\\
\emph{\indent SSS Process.} In this process, each sub-secret inside each group is distributed to group members. The final result is that each server stores both an original secret share of secret s and a sub-share of generated sub-secret $s_{i}$, which means the capacity will be doubled to $2n$. Hence, the additional capacity turns out to be n.\\
\subsubsection{Threshold}As shown in the context, in our model the maintenance of threshold is conditional, depending on the security of repairing data. When this condition is not met, the scenario turns out to be a perfect access structure rather than a threshold scheme.
\subsubsection{Perfect Security}Both of them achieve perfect security.
\subsubsection{Repairing Locality}Regenerating codes can not achieve repairing locality. In our model, the redundancy distribution protocol considers a trade-off between repairing locality and security. Only 1 extra share is banished with a little impact on locality.
\subsubsection{Computational Complexity.} We have pointed out in the previous discussion that one advantage of the grouping scenario is that it has lower computational complexity in the coding and decoding process. For example in the repairing function generation process, what we need is nothing but solving liner equations, which can be solved with a complexity between $O(n^{2})$ and $O(n^{3})$, where $n$ is the total number of shares in a $(k,n)$ SS scheme.\\
\indent Consider the $n$ participants who are divided into $m$ groups. Suppose there are $\gamma$ members in each group such that
\begin{equation}
n=m\gamma
\end{equation}

Assume that:
\begin{enumerate}
\item In each group, at most 1 failure happens at any given moment.
\item In each group, the probability of server failure is $q$.
\end{enumerate}

Then, the computational complexity turns out to be
\begin{equation}
O(qm\gamma^{3})=O(qn\gamma^{2})=O(\gamma^{2}n)
\end{equation}

When $\gamma$ is determined and small, the complexity turns out to be $O(n)$. As we can see, from this aspect, $\gamma$ should be made as low as possible to achieve lower computational complexity.\\
\indent However, the upper boundary of the SS threshold turns out to be $\gamma$ so it is better not to be too small or the security level will be low. In this paper the suggested value for $\gamma$ is 4.\\
\indent The results are listed in TABLE \uppercase\expandafter{\romannumeral1}.
\begin{table}[!hbp]
\center
\caption{Performance Comparison with Regenerating Codes}
\begin{tabular}{|c|c|c|}
\hline
 & Regenerating Codes & Our Model\\
\hline
Storage & - & additional\\
\hline
Perfect Security & yes & yes\\
\hline
Threshold & yes & conditional\\
\hline
Repairing Locality & no & feasible\\
\hline
Computational Complexity & high & low\\
\hline
\end{tabular}
\end{table} 

\section{Conclusion}This paper aims at share repairing issues in a SS scheme. The studies in the literature will either lose threshold property of the original SS scheme or their repairing degrees are high. In this paper a grouped repairing mechanism based on repairing function is proposed, which splits secret shares into disjoint groups to achieve repairing locality. Repairing redundancies are generated to keep repairing informations. Furthermore, IDA is proposed to make the repairing process almost restricted inside groups so locality is achieved. Finally, we indicate in our scheme the maintenance of threshold property is conditional. In the worst case, with the loss of security of repairing data, the original threshold SS scheme turns out to be an access structure. Still, the secure level is equivalent to existing scenarios. In any case, both locality and perfect security are still remained.


\begin{thebibliography}{1}
\bibitem[1]{1}Adi Shamir, How to share a secret, Communications of the ACM, 22(11):612-613, Nov 1979.
\bibitem[2]{2}Blakley G R. Safeguarding cryptographic keys[C]// afips. IEEE Computer Society, 1979:313.
\bibitem[3]{3}Pan D, Kuang X H, Xi-Cheng L U. A Non-Interactive Protocol for Member Expansion in a Secret Sharing Scheme[J]. Journal of Software, 2005, 16(1):116-120.
\bibitem[4]{4}Xu J F, Cui G H, Cheng Q, et al. Cryptanalysis of a non-interactive protocol for member expansion in a secret sharing scheme[J]. Journal on Communications, 2009, 30(10):118-123.
\bibitem[5]{5}Nojoumian M, Stinson D R, Grainger M. Unconditionally secure social secret sharing scheme[J]. Iet Information Security, 2010, 4(4):202-211.
\bibitem[6]{6}Wu Y, Li D, Wang F. Secret Sharing Member Expansion Protocol Based on ECC[J]. Open Cybernetics \& Systemics Journal, 2015, 8(1):248-253.
\bibitem[7]{7}Herzberg A, Jarecki A. Proactive Secret Sharing Or: How to Cope With Perpetual Leakage[J]. Advances in Cryptology-Crypto'95, 1998, 963:339-352.
\bibitem[8]{8}Yu J, Kong F, Hao R. Publicly Verifiable Secret Sharing with Enrollment Ability[C]// Eighth Acis International Conference on Software Engineering, Artificial Intelligence, Networking, and Parallel/distributed Computing. IEEE Xplore, 2007:194-199.
\bibitem[9]{9}Chun-Gen X U, Yang Y J. Protocol for Member Expansion in Publicly Verifiable Secret Sharing Scheme[J]. Journal of Nanjing University of Science \& Technology, 2009.
\bibitem[10]{10}Schoenmakers B. A Simple Publicly Verifiable Secret Sharing Scheme and Its Application to Electronic Voting[M]// Advances in Cryptology ¡ª CRYPTO¡¯ 99. Springer Berlin Heidelberg, 1999:148--164.
\bibitem[11]{11}Saxena N, Tsudik G, Yi J H. Efficient node admission for short-lived mobile ad hoc networks[C]// IEEE International Conference on Network Protocols. IEEE Xplore, 2005:10 pp.
\bibitem[12]{12}Yue B I. Protocol for member expansion in vector space secret sharing[J]. Computer Engineering \& Applications, 2011, 47(16):74-76.
\bibitem[13]{13}AG Dimakis, K Ramchandran, Y Wu, C Suh, Network Coding for Distributed Storage Systems, IEEE TRANSACTIONS ON INFORMATION THEORY, VOL. 56, NO. 9, 2010
\bibitem[14]{14}Xuan Guang, Jiyong Lu and Fang-Wei Fu, Repairable Threshold Secret Sharing Schemes, arXiv:1410.7190v2, 2015.
\bibitem[15]{15}Ankit Singh Rawat, O. Ozan Koyluoglu and Sriram Vishwanath, Centralized repair of multiple node failures, 2016 IEEE International Symposium on Information Theory (ISIT), Pages: 1003 - 1007, DOI: 10.1109/ISIT.2016.7541450, 2016.
\bibitem[16]{16}Papailiopoulos, Dimitris S., A. G. Dimakis. "Locally Repairable Codes." IEEE Transactions on Information Theory 60.10(2012):5843-5855.
\bibitem[17]{17}Agarwal, A, and A. Mazumdar. Security in locally repairable storage. Information Theory Workshop IEEE, 2015:1-5.
\bibitem[18]{18}H. Krawczyk , Secret Sharing Made Short, Advances in Cryptology ¡ª Crypto¡¯ 773(1988):136-146, 1988.
\bibitem[19]{19}Hayashi D., et al, Design and implementation of autonomous distributed secret sharing storage system Communications, Apcc 2003. the, Asia-Pacific Conference on 2003:57-60 Vol.1, 2003.
\bibitem[20]{20}Asmuth C, Bloom J. A modular approach to key safeguarding[J]. Information Theory IEEE Transactions on, 1983, 29(2):208-210.
\bibitem[21]{21}Karnin E, Greene J, Hellman M. On secret sharing systems[J]. IEEE Transactions on Information Theory, 1983, IT-29(1):35-41.
\bibitem[22]{22}Blakley, G. R, and C. Meadows, Security of ramp schemes. Advances in Cryptology, Proceedings of CRYPTO '84, Santa Barbara, California, USA, August 19-22, 1984, Proceedings DBLP, 1984:242-268.
\bibitem[23]{23}Kurihara, M, and H. Kuwakado. Secret sharing schemes based on minimum bandwidth regenerating codes. International Symposium on Information Theory and ITS Applications 2012:255-259.
\bibitem[24]{24}Herzberg A, Jarecki A. Proactive Secret Sharing Or: How to Cope With Perpetual Leakage[J]. Advances in Cryptology-Crypto'95, 1998, 963:339-352.
\bibitem[25]{25}K.V. Rashmi, N.B. Shah, P.V. Kumar, Optimal Exact-Regenerating Codes for Distributed Storage at the MSR and MBR Points via a Product-Matrix Construction, IEEE Transactions on Information Theory 57(57):5227-5239, 2011
\bibitem[26]{26}S. Pawar, S. El Rouayheb, K. Ramchandran, Securing
Dynamic Distributed Storage Systems Against Eavesdropping and Adversarial Attacks, IEEE Trans. Information Theory, pp.6734¨C6753, 2011.
\bibitem[27]{27}Papailiopoulos D S, Dimakis A G. Locally Repairable Codes[J]. IEEE Transactions on Information Theory, 2012, 60(10):5843-5855.
\bibitem[28]{28}Shahabinejad M, Khabbazian M, Ardakani M. An Efficient Binary Locally Repairable Code for Hadoop Distributed File System[J]. IEEE Communications Letters, 2014, 18(8):1287-1290.
\bibitem[29]{29}Goparaju S, Calderbank R. Binary cyclic codes that are locally repairable[M], 2014.
\bibitem[30]{30}Silberstein N, Rawat A S, Koyluoglu O O, et al. Optimal locally repairable codes via rank-metric codes[J]. Mathematics, 2013, 31(6):1819-1823.
\bibitem[31]{31}Zeh A, Yaakobi E. Optimal linear and cyclic locally repairable codes over small fields[C]// Information Theory Workshop. IEEE, 2015:1-5.
\end{thebibliography}
\end{document}